\documentclass[aps,twocolumn,superscriptaddress]{revtex4}
\usepackage[dvips]{graphicx}
\usepackage{amsfonts} 
\usepackage{amsmath} 
\usepackage{latexsym} 
\usepackage{amssymb}
\usepackage{eurosym}
\usepackage{color}

\newcommand{\nonqip}[1]{#1}
\newcommand{\qip}[1]{}

\newcommand{\abs}[1]{\left|#1\right|}

\newcommand{\ket}[1]{| #1 \rangle}
\newcommand{\bra}[1]{\langle #1 |}

\newcommand{\trace}{\textrm{Tr}}

\newcommand{\Z}{\mathbb{Z}}
\newcommand{\N}{\mathbb{N}}

\newcommand{\joliC}{\mathcal{C}}
\newcommand{\joliH}{\mathcal{H}}

\newcommand{\joliN}{\mathcal{N}}
\newcommand{\joliE}{\mathcal{E}}
\newcommand{\joliV}{\mathcal{V}}
\newcommand{\pa}[1]{\left(#1\right)}
\newcommand{\acco}[1]{\left\{#1\right\}}

\newtheorem{Def}{Definition}
\newtheorem{Th}{Theorem}

\newtheorem{Cor}{Corollary}
\newtheorem{Pro}{Proposition}

\begin{document}

\title{\begin{center}
Unitarity plus causality implies localizability
\end{center}}

\begin{abstract}
We consider a graph with a single quantum system at each node. The entire compound system evolves in discrete time steps by iterating a global evolution $U$. We require that this global evolution $U$ be unitary, in accordance with quantum theory, and that this global evolution $U$ be causal, in accordance with special relativity. By causal we mean that information can only ever be transmitted at a bounded speed, the speed bound being quite naturally that of one edge of the underlying graph per iteration of $U$. We show that under these conditions the operator $U$ can be implemented locally; i.e. it can be put into the form of a quantum circuit made up with more elementary operators --- each acting solely upon neighbouring nodes. We take quantum cellular automata as an example application of this representation theorem: this analysis bridges the gap between the axiomatic and the constructive approaches to defining QCA. \qip{{ The longer version of this extended abstract is available at \cite{ArrighiUCAUSAL}.}}
\end{abstract}


\author{Pablo Arrighi}
\email{pablo.arrighi@imag.fr}
\affiliation{Universit\'e de Grenoble, Laboratoire LIG, 220 rue de la Chimie,
 38400 SMH, France}

\author{Vincent Nesme}
\email{vincent.nesme@itp.uni-hannover.de} 

\author{Reinhard Werner}
\email{reinhard.werner@itp.uni-hannover.de} 
\affiliation{Leibniz Universit\"at Hannover,\\ 
Institut f\"ur theoretische Physik, Appelstr. 2, 30167 Hannover, Germany. }

\maketitle

\section{Motivations}

A physical system is described in quantum theory by a state vectors $\ket{\psi}$ (a unit vector in a Hilbert space $\mathcal{H}$), and evolves from time $t$ to time $t'$ according to a unitary operator $U$. The definition of a unitary operator is made in general terms: induce an bijection between two orthonormal bases ($U^{\dagger}U=U^{\dagger}U=\mathbb{I}$). But what if we want a more hands-on, operational description of a unitary operator? In finite dimension we know that they can be spectrally decomposed ($U=\sum_x e^{i\lambda_x}\ket{\phi_x}\bra{\phi_x}$), but also that they can be approximated up to arbitrary precision by a circuit composed of the universal quantum gates $H$, $Phase$, $C-Not$ --- via the Solovay-Kitaev theorem. However in infinite dimensions spectral theory becomes quite complicated, and nothing tells us whether the operator can be expressed as a quantum circuit. Often it can be difficult to provide an operational description of unitary operators over infinite dimensional spaces; in  that sense they remain abstract mathematical objects.\\
Usually the infinite number of degrees of freedom arises from a position degree of freedom, i.e. when space comes into the picture. The canonical example is that of the wave function of a particle on a line. But physics then tells us something else about the evolution, namely that if the particle is well-localized within a region $R$ at time $t'$, it was not to be found outside the region $R\pm c(t'-t)$ at time $t$. This is a case of causality, and what causality says in general is that if we distinguish different ``places'', some of them close to one another, some of them distant, and if the interval $(t'-t)$ is sufficiently small, then the state associated to some place $x$ at time $t'$ should only depend upon the state associated to the neighbours of $x$ at time $t$.\\
\qip{Let us make this notion a bit more formal.} A composite physical system $AB$ is described in quantum theory by a state vector $\ket{\psi}^{AB}$ (a unit vector in $\mathcal{H}^{A}\otimes\mathcal{H}^{B}$), which in general cannot be decomposed into state vectors $\ket{\psi}^{A}/\ket{\psi}^{B}$ associated to subsystems $A/B$ --- due of course to entanglement. In order to still be able to speak of ``state associated to some place'' we must switch to the well established formalism of states and partial traces. So now a composite physical system is described by a state $\rho^{AB}$ (i.e. a unit trace positive operator over $\mathcal{H}^A\otimes\mathcal{H}^B$), and the state associated to place $A$ is $\rho|_{A}=\trace_B(\rho)$. 
\qip{Moreover so far we have been speaking about physical systems divided up in ``places, some of them close to one another, some of them distant''. As we seek to capture this idea in the most general, and yet simple and formal manner, we shall identify those ``places'' with the nodes of an arbitrary graph, say that two nodes are ``close'' whenever they are related by an edge, and say that there is a physical system sitting at each of these nodes.}
And hence we can express causality by saying that if $\rho$ taken to $\rho'$ \nonqip{over a short enough period of time}\qip{in one time step}, then $\rho'|_x$ should be a function of $\rho|_{\joliN_x}$, with $\joliN_x$ designating the neighbours of $x$.\\
In this paper we \nonqip{will} study unitary causal operators for they own sake, and we \nonqip{will} show that they have a lot of structure. For instance the inverse of a unitary causal operator is also a unitary causal operator. More importantly, we \nonqip{will} show that they decompose into a product of local operators, i.e. operators which act solely upon neighbourhoods. This structure theorem is a general representation theorem for unitary causal operators, which yields an operational description of them --- just like the two aforementioned representation theorems did for finite-dimensional unitary operators.

This structure theorem follows a tradition of postulating causality over a global dynamics and then showing that the dynamics can be implemented locally --- which is the difficult direction to go to of course (mainly due to entanglement), the converse direction being always trivial. It provides a general answer to this question under the unitarity and discreteness hypotheses, encompassing (up to some details) the previous results on this issue by Beckman et Al. \cite{Beckman} for two systems, Schumacher and Westmoreland \cite{SchumacherWestmoreland} for three systems, and Schumacher and Werner for a line of translation-invariant one-dimensional systems \cite{SchumacherWerner}, i.e. Quantum cellular automata (QCA). \nonqip{As regards QCA more specifically, a corollary of our results is that the axiomatic definition of $n$-dimensional QCA proposed in \cite{SchumacherWerner} does admit an operational, Block-structured QCA description after all --- and conversely that these seemingly specific Block-structured QCA \cite{PerezCheung} are in fact general instances axiomatic definition. }

\nonqip{
We start with basic definitions and properties (Sections \ref{secdefs} and \ref{secprops}) before we prove our main result in Section \ref{blocks}. We then discuss related works mainly in quantum cellular automata, where we take our inspirations from, and discuss the more general perspectives (Sections \ref{qca} and \ref{pers}).
}

\nonqip{
\section{Definitions}
\label{secdefs}

\noindent So far in this intuitive motivation towards the concept of unitary causal operator we have been speaking about ``places, some of them close to one another, some of them distant''. As we seek to capture this idea in the most general, and yet simple and formal manner, we shall identify those ``places'' with the nodes of an arbitrary graph, and say that two nodes are ``close'' whenever they are related by an edge. We first need to make rigorous the idea of a graph, with a quantum system at each node.
\begin{Def}[quantum labeled graph]~\label{def:qlg}\\ 
A quantum labeled graph (QLG) is a tuple $\Gamma=(\joliV,\joliE,\joliH)$ with:\\
- $\joliV$, the nodes (a countable set);\\
- $\joliE$, the edges (a subset of $\joliV\times \joliV$);\\
- $\joliH$, the labels (a countable set of Hilbert spaces).\\
We denote by $\joliN_x=\{ y\,|\,(x,y)\in \joliE \}$ the set of direct neighbours of the node $x$, with $x$ an integer ranging over $\joliV$.
\end{Def}
To each node $x$ there is an associated alphabet $\Sigma^x$ and hence an Hilbert space $\joliH^x=\joliH_{\Sigma^x}$. Now say the graph is infinite. The difficulty here is that as we have mentioned an infinite tensor product Hilbert spaces ``$\bigotimes_\mathbb{N} \joliH^x$'' is in general not a Hilbert space, so we must take the following detour:
\begin{Def}[(finite) configurations]\label{def:Cfbis}~\\ 
A (finite) configuration $c$ of a QLG $\Gamma=(\joliV,\joliE,\joliH)$ is a function $c: \mathbb{N} \longrightarrow \mathbb{N}$, with $x\longmapsto c(x)=c_x$, such that:\\
- $c_x$ belongs to $\Sigma^x$;\\
- the set $\{x\,|\,c_x \neq q\}$ is finite.\\ 
The set of all finite configurations of a QLG be denoted $\joliC_f$ again.
\end{Def}
The idea is that finite configurations are the basic states of the quantum systems labelling the graph. The following definition works because $\joliC_f$ is countable:
\begin{Def}[superpositions of configurations]~\label{def:HCfbis}\\ 
We define $\joliH_{\joliC_f}$ be the Hilbert space of configurations of a QLG $\Gamma=(\joliV,\joliE,\joliH)$, as follows: to each finite configurations $c$ is associated a unit vector $\ket{c}$, such that the family $\pa{\ket{c}}_{c\in\joliC_f}$ is an orthonormal basis of $\joliH_{\joliC_f}$. A \emph{state vector} is a unit vector $\ket{\psi}$ in $\joliH_{\joliC_f}$. A \emph{state} is a trace-one positive operator $\rho$ over $\joliH_{\joliC_f}$. 
\end{Def}
Note that $\joliH_{\joliC_f}$ is entirely defined by the set of Hilbert spaces $\joliH=(\joliH^x)$. From now on we will write $\joliH$ instead of $\joliH_{\joliC_f}$. Note also that the state $\rho$ captures the state of the entire compound system, whereas $\rho|_x$ stands for the state which labels node $x$ of the graph, where we introduce the notation $A|_{\mathcal{S}}$ for the matrix $\trace_{\textrm{All but the systems in $\mathcal{S}$}}(A)$. 
\begin{Def}[Causality]~\label{def:causality}\\ 
A linear operator $U:\joliH\longrightarrow\joliH$ is said to be \emph{causal} with respect to a quantum labeled graph $\Gamma$ if and only if for any $\rho,\rho'$ two states over $\joliH$, and for any $x\in\Z$, we have
\begin{align}
\rho|_{\joliN_x}=\rho'|_{\joliN_x}\quad \Rightarrow U(\rho)U^\dagger|_{x}=U(\rho')U^\dagger|_{x}. \label{eq:causality}
\end{align}
\end{Def}
In other words: to know the state of node number $x$, we only need to know the neighbouring of nodes $\joliN_x$. Unitarity is as usual:
\begin{Def}[Unitarity]~\label{def:unitarity}\\ 
A linear operator $U:\joliH\longrightarrow\joliH$ is \emph{unitary} if and only if $\{U\ket{c}\,|\,c\in\mathcal{C}_f\}$ is an orthonormal basis of $\joliH_{\joliC_f}.$
\end{Def}
Hence we have defined the main object of our discourse: unitary causal operators. This concept of unitary causal operator generalizes the two-systems definition by Beckman, Gottesman, Nielsen, Preskill \cite{Beckman} and the three-systems definition by Schumacher and Westmoreland \cite{SchumacherWestmoreland}.

\section{Properties}\label{secprops}

Let us begin by proving some fundamental facts about unitary causal operators, which may also be regarded as alternative formulations of causality. \noindent Proposition \ref{prop:dual} expresses causality in the Heisenberg picture, as a condition on the evolution of observables. Whenever we say that a linear operator $A$ is localized upon a region $R$, we mean that $A$ is of the form $A_{R}\otimes\mathbb{I}_{\joliV\setminus R}$, i.e. it is the identity over anything that lies outside of $R$. Morally, $A$ is an observable in the following result.
\begin{Pro}[Dual causality]~\\ \label{prop:dual}
Let $U$ be a  causal linear operator with respect to a quantum labeled graph $\Gamma$. This is equivalent to saying that for every operator $A$ localized upon node $x$, then $U^\dagger AU$ is localized upon the nodes in $\joliN_x$.
\end{Pro}
\textbf{Proof.} 
$[\Rightarrow]$. Suppose causality and let $A$ be an operator 
localized upon node $x$. For every states $\rho$ and $\rho'$ such that 
$\rho|_{\joliN_x}=\rho'|_{\joliN_x}$, we have $\pa{U\rho U^\dagger}|_x=\pa{U\rho' U^\dagger}|_x$ and hence $\trace\pa{AU\rho U^\dagger}=\trace\pa{AU\rho' U^\dagger}$. We thus 
get $\trace\pa{U^\dagger AU\rho}=\trace\pa{U^\dagger AU\rho'}$. Since 
this equality holds for every $\rho$ and $\rho'$ such that 
$\rho|_{\joliN_x}=\rho'|_{\joliN_x}$, what we are saying is that the  $U^\dagger AU$
does not discriminate differences between $\rho$ and $\rho'$ whenever 
they lie outside of ${\joliN_x}$. In other words $U^\dagger AU$ is localized 
on the nodes in ${\joliN_x}$.\\
$[\Leftarrow]$. Suppose dual causality and $\rho|_{\joliN_x}=\rho'|_{\joliN_x}$. 
Then, for every operator $B$ localized upon the nodes in $\joliN_x$, 
$\trace\pa{B\rho}=\trace\pa{B\rho'}$, and so for every operator $A$ localized upon node $x$, we get: 
$\trace\pa{AU\rho U^\dagger}=\trace\pa{U^\dagger 
AU\rho}=\trace\pa{U^\dagger AU\rho'}
=\trace\pa{AU\rho'U^\dagger}.$
This entails $\pa{U\rho U^\dagger}|_x=\pa{U\rho' U^\dagger}|_x$.\hfill $\Box$
Proposition \ref{prop:inv} expresses causality in terms of the inverse of the unitary causal operator $U$. Whenever we speak about the transpose of a quantum labeled graph $\Gamma$, we mean as usual the quantum labeled graph $\Gamma^T$ which is obtained just by changing the direction of the edges. The neighbours of $x$ in $\Gamma^T$ are designated by $\joliN^T_x$.\\
\begin{Pro}[Inverse causality]~\\ \label{prop:inv}
Let $U$ be a  causal linear operator with respect to a quantum labeled graph $\Gamma$. Then $U^\dagger$ is a causal operator with respect to the transposed quantum labeled graph $\Gamma^T$.
\end{Pro}
\textbf{Proof.} 
Suppose causality, let $A$ be an operator localized upon node $x$, and choose $M$ an operator localized upon a node $y$ which does not lie in $\joliN^T_x$. That way $x$ does not belong to $\joliN_y$. But according to Proposition \ref{prop:dual} we know that $U^{\dagger}MU$ is localized upon $\joliN_y$, and hence $U^{\dagger}MU$ commutes with $A$. Now $A\mapsto UAU^\dagger$ is a morphism because $AB\mapsto UAU^{\dagger}UBU^{\dagger}=UABU^{\dagger}$, and so via this morphism we can also say that $UU^{\dagger}MUU^{\dagger}=M$ commutes with $UAU^{\dagger}$. An since $M$ can be chosen amongst to full matrix algebra $M_d(\mathbb{C})$ of the node $y$, this entails that $UAU^{\dagger}$ must be the identity upon this node. The same can be said of any node outside $\joliN^T_x$. So $UAU^{\dagger}$ is localized upon $\joliN^T_x$ and we can conclude our proof via Proposition \ref{prop:dual}. \hfill $\Box$\\
We will use both these propositions in order to establish our representation main theorem.

}

\nonqip{
\section{Representation}\label{blocks}

We will now show that unitary causal operators are implementable locally; i.e. that they can be put into the form of a quantum circuit made up with more elementary operators -- each acting solely upon neighbouring nodes.
\begin{Th}[Local representation]~\\ \label{th:locrep}
Let $U$ be a unitary causal operator with respect to a quantum labeled graph $\Gamma=(\joliV, \joliE, \joliH)$. Then there exists $D$, $(K_x)$, $E$, and $\ket{\phi}$ such that for all $\ket{\psi}$,
$$(\bigotimes D)(\prod K_x)(\bigotimes E)\ket{\psi}=\ket{\phi}\otimes U\ket{\psi}$$
where:
\begin{itemize}
\item $(K_x)$ is a collection of commuting unitary operators localized upon each neighbourhood $\joliN^T_x$;\\
\item $D^{\dagger}, E$ are two isometric operators localized upon each node $x$, and whose actions depend only on $\dim(\joliH^x)$.\\
\end{itemize}
Moreover:\\
\begin{itemize}
\item If the $(\joliH^x)$ are all of finite dimensions, then the $(K_x), D^{\dagger}$ and $E$ are finite dimensional operators;\\
\item If $U(\bigotimes \ket{q})=(\bigotimes \ket{q})$, then $\ket{\phi}=(\bigotimes \ket{q})$;\\
\item If the $(\joliH^x)$ are all of infinite dimensions and $U(\bigotimes \ket{q})=(\bigotimes \ket{q})$, then we can choose to just have $(\bigotimes D)(\bigotimes K_x)(\bigotimes E)=U\ket{\psi}$ where $D$ and $E$ are also unitary.
\end{itemize}
\end{Th} 
\textbf{Proof.} 
$[\textrm{Encoding}].$ The action of $E$ upon node $x$ is just to add an ancilla, i.e. $E\ket{\psi_x}=\ket{q}\otimes\ket{\psi_x}$. Hence if $\dim(\joliH^x)$ is finite then $E:\joliH^x\longrightarrow \joliH^x \otimes \joliH^x$ and $E$ is an isometry, whereas if $\joliH^x$ is of infinite countable dimension then we can use any bijection from $\N\times\N$ to $\N$ so that $E:\joliH^x\longrightarrow \joliH^x$ and $E$ is unitary. \\
$[\textrm{Product states}].$ Let us consider $\ket{\psi}\in\joliH$ having the form of a product state, i.e. so that $\ket{\psi}=\bigotimes \ket{\psi_x}$. We will show that $(\bigotimes D^\dagger)(\bigotimes K_x)(\bigotimes E)\ket{\psi}=U\ket{\psi}$, and then by linearity the result will be proved for entangled states also. This is because in general any state vector $\ket{\phi}$ can be written as a sum of $\ket{\phi^i}$, where each $\ket{\phi^i}$ is a product state $\ket{\phi^i}=\bigotimes \ket{\phi^i_x}$. Below we again use this form for $\ket{\phi}=U^{\dagger}(\bigotimes \ket{q})$. \\
$[\textrm{Two tapes}].$
So $E$ takes $\ket{\psi}$ into $(\bigotimes \ket{q})\otimes (\bigotimes \ket{\psi_{x}})$. Now since $(\bigotimes \ket{q})=UU^\dagger (\bigotimes \ket{q}) =U\ket{\phi}$ we rewrite $E\ket{\psi}$ as:
$$\sum_i U(\bigotimes \ket{\phi^i_x})\otimes (\bigotimes \ket{\psi_x})$$
So initially our QLG has got two ``tapes'', one which we call the ``computed tape'' holding state $U(\bigotimes \ket{\phi})$, and one which we call the ``uncomputed tape'' holding state $(\bigotimes \ket{\psi_x}).$\\
$[\textrm{Changing factors}].$ Now the idea is that the $K_x$ will let us pass pieces of the uncomputed tape to the computed tape. Namely we want $K_x E\ket{\psi}$ is equal to:
$$\sum_i U\big(\ket{\psi_{x}}\otimes\bigotimes_{\joliV\setminus\{x\}} \ket{\phi^i_y}\big)\otimes \big(\ket{\phi^i_x}\otimes\bigotimes_{\joliV\setminus\{x\}} \ket{\psi_{y}}\big).$$
Let us simply take $K_x=U\, Swap_x\, U^\dagger$, meaning that we simply uncompute the computed tape, swap $\ket{\psi_x}$ for $\ket{q}$, and then compute it back. Clearly this does the job but does seem wrong, because it looks as though we are acting over the entire graph and not just $\joliN_x$. Yet this naive choice is actually the right one. Indeed since $U$ is unitary causal with respect to $\Gamma$, then so is $U^\dagger$ with respect to $\Gamma^T$, by Proposition \ref{prop:inv}. And now since $U^\dagger$ is unitary causal over $\Gamma^T$, $U\, Swap_x\, U^\dagger$ must be localized upon $\joliN^T_x$, by virtue of Proposition \ref{prop:dual}. Note that the $(K_x)$ commute with one another just because the $(Swap_x)$ commute with one another and $A\mapsto UAU^\dagger$ is a morphism.\\
$[\textrm{Decoding}].$ Of course we can reiterate this process until we get
$$ \sum_i U (\bigotimes \ket{\psi_{x}})\otimes (\bigotimes \ket{\phi^i_x})$$
which is just $U\ket{\psi}\otimes \ket{\phi}$. Now we just need to swap the computed and uncomputed tapes to get $(\ket{\phi}\otimes U\ket{\psi}$. (In situations where $U^{\dagger}(\bigotimes \ket{q})=\ket{\phi}$ is known and turns out to be a product state $\bigotimes \ket{\phi_x}$, then $D$ can also locally undo the $\ket{\phi}$ so as to get $(\bigotimes \ket{q})\otimes U\ket{\psi}=E U \ket{\psi}$. This is the case for instance in the standard situation when $U(\bigotimes \ket{q})=(\bigotimes \ket{q})$. If on top of that $E$ was a unitary, $D$ can also apply $E^\dagger$ and give back $U \ket{\psi}$.) \hfill $\Box$\\

\begin{Cor}[Circuit representation]~\\ \label{cor:circrep}
Let $U$ be a unitary causal linear operator with respect to a quantum labeled graph $\Gamma=(\joliV, \joliE, \joliH)$. Then $U$ can be expressed as a circuit of quantum operations each localized upon a neighbourhood $\joliN^T_x$, and having depth less or equal to $\deg(\Gamma)^2+2$.
\end{Cor}
\textbf{Proof.} By inspection of the proofs of Theorem \ref{th:locrep} and using the following remarks. Since each $K_x$ is localized upon $\joliN^T_x$, many of them can be done in parallel, namely whenever the corresponding neighbourhoods do not intersect. The question of how much can be done in parallel, i.e. how many layers of circuit are necessary, is equivalent to the $L(1,1)$-labeling problem for graphs, namely we want to colour the graph so that no neighbours nor next-neighbours have the same colours. This is known to require at most $\deg(\Gamma)^2$ colours \cite{ChangMEHDI}. The plus two is for $E$ and $D$. \hfill $\Box$\\
} 
The study of unitary causal operators has older origins than the rise quantum information processing, for similar questions are clearly coming up in axiomatic/algebraic quantum field theories\nonqip{ \cite{Buchholz}} --- as argued also in the papers which treat the two-systems and three-systems cases of this theorem \cite{Beckman, SchumacherWestmoreland}. The main difference in approach seems to be that AQFT caters looks at continuous time and space. The authors are not aware, however, of a result akin to \qip{this structure theorem}\nonqip{Theorem \ref{th:locrep}} in AQFT, which would let us structure the dynamics of the system in such a meaningful, operational manner. But since our initial motivation was the study of QCA --- let us see what this result has to say about them.

\section{Consequences and related results in QCA}\label{qca}

Cellular automata (CA) as introduced by Von Neumann \cite{Neumann}, consist of an array of identical cells, each of which may take one in a finite number of possible states. The whole array evolves in discrete time steps by iterating a function $G$. Moreover this global evolution $G$ is shift-invariant (it acts everywhere in the same way) and local (information cannot be transmitted faster than some fixed number of cells per time step). Because this is a physics-like model of computation, Feynman suggested right from the birth of quantum computation \cite{FeynmanQCA} that one should look into quantizing this model (for two reasons: first because in CA computation occurs without extraneous control, hence this gets rid of a source of decoherence; second because they are a good framework to study quantum simulation of a quantum system). 
Quantum cellular automata were always going to be unitary operators $G$ over arrays of finite-dimensional systems, together with shift-invariance (``the laws of physics are everywhere the same'') and causality (``there can be no instantaneous long-range communication''). At an informal level this concept has been around for almost twenty years, unchallenged but yet somewhat impractical --- in the sense that there was no proper axiomatization and nor a generic operational description of them. And so it was not really know what these things actually looked like. As Gruska puts it in one of the very first textbook on quantum computation \cite{Gruska}: ``A suitable definition of two- and more-dimensional quantum cellular automata is an untrivial matter.''\\
This situation has led to several competing definitions of QCA, each one attempting to tame the structure of the unitary operator $G$ in its own manner. Let us briefly look at the three main approaches towards defining QCA\nonqip{ --- excluding by lack of space those works which are more concerned with quantum walks \cite{GrossingZeilinger, MeyerQLGI}, quantum simulations \cite{BoghosianTaylor1, LoveBoghosian}, or implementation models \cite{VollbrechtCirac, FitzsimonsTwamley}}. Historically the first approach \cite{Watrous, DurrWell, ArrighiMFCS} was recently shown to break causality \cite{ArrighiLATA}, and so it seems we must abandon this definition. The second approach \cite{SchumacherWerner} is the axiomatic one, it provides a rigorous axiomatics for quantum cellular automata, which we can rephrase in the vocabulary of this paper as \nonqip{follows:
\begin{Def}[QCA]~\label{def:qca}\\ A $n$-dimensional quantum cellular 
automaton (QCA) is an shift-invariant unitary causal operator $G:\joliH\longrightarrow\joliH$
over a QLG $\Gamma=(\joliV,\joliE,\joliH)$ with:\\
- $\joliV=\mathbb{Z}^n$ i.e. the nodes form a grid;\\
- $\joliE=\{x,x+z\;|\;x\in\mathbb{Z}^n\,\wedge\,z\in\{0,1\}^n\}$ i.e. radius half;\\
- $\joliH=(\joliH_\Sigma)$ i.e. all cells are of a given finite dimension $d$.
\end{Def}}\qip{just being translation-invariant unitary causal operators, over the grid.}
This paper \cite{SchumacherWerner} then derives a Block structure in a very general fashion. Unfortunately the proof is flawed in $n$-dimensions \cite{ArrighiLATA}. Fortunately the result contained in this paper entails another Block representation of $n$-dimensional QCA\qip{.}\nonqip{:}
\nonqip{
\begin{Th}[$n$-dimensional QCA]\label{th:multilayers}~\\
Let $G$ be an $n$-dimensional QCA with alphabet $\Sigma$. Let $E$ be an isometry from $\joliH_\Sigma\to\joliH_\Sigma\otimes\joliH_\Sigma$ such that $E\ket{\psi_x}=\ket{q}\otimes\ket{\psi_x}$. This mapping can be obviously extended to whole configurations, yielding a mapping $E:\joliH_{C^{\Sigma}_f}\to\joliH_{C^{\Sigma^2}_f}$. Then there exists a $n$-dimensional QCA $H$ on alphabet $\Sigma^2$, such that $HE=EG$, and $H$ admits an $2^n$-layer block representation. Moreover $H$ is of the form 
\begin{align}
H=(\bigotimes S)(\prod K_x) \label{eq:luqca}
\end{align}
where:
\begin{itemize}
\item $(K_x)$ is a collection of commuting unitary operators all identical up to shift, each localized upon each neighbourhood $\joliN_x$;\\
\item $S$ is the swap gate over $\joliH_\Sigma\otimes\joliH_\Sigma$, hence localized upon each node $x$.\\
\end{itemize}
\end{Th}
\textbf{Proof.} By inspection of the proof of Theorem \ref{th:locrep} and using the following remarks. At each $x=(i_1,\ldots,i_n)$ step,  $K_x=K_{i_1\ldots i_n}$ is local to cells $\{i_1,i_1+1\}\times\ldots\times\{i_n,i_n+1\}$, uncomputed and computed tapes alike. Namely, whenever $(i_1,\ldots,i_n)$ and $(j_1,\ldots,j_n)$ are such that for every $k\in\left\{1,\ldots,n\right\}$, $\abs{i_k-j_k}>1$, then $K_{i_1\ldots i_n}$ and $K_{j_1\ldots j_n}$ can be performed in parallel. So we can first apply simultaneously all the $K_{i_1,\ldots, i_n}$'s where the $i_k$'s are even. Then, as each element $x=(x_1,\ldots,x_n)$ can be written in a unique way as the sum of $y$ with even coordinates and $z\in\acco{0,1}^n$, we need $\abs{\acco{0,1}^n}=2^n$ layers to apply all of the $K_{i_1,\ldots,i_n}$'s. Moreover by shift-invariance these $K_{i_1\ldots i_n}$'s are just shifted version of the same $K$, so that each layer is just tiling of the space by a finite-dimensional unitary $K$.
\hfill $\Box$\\
(Notice that even the shift QCA can be implemented as a quantum circuit according to this theorem, but of course this is at the price of introducing ancilla. This question of the special role of the shift and its implementability is emphasized in \cite{PerezCheung} and carefully analysed in \cite{GrossNesmeVogtsWerner}.)\\
}
The third approach to define QCA is to give them an a priori hands-on, operational description of a particular form. Several works have followed this route for instance \cite{VanDam, BrennenWilliams, NagajWocjan, Raussendorf}, but amongst them \cite{PerezCheung} stands out at this stage as it just directly posits, after some interesting arguments, that their evolutions takes a form akin to the one described in given by \qip{our structure theorem}\nonqip{Theorem \ref{th:multilayers}}.\\
Here we have demonstrated that starting just from the axiomatic definition of QCA as in \cite{SchumacherWerner} and \cite{ArrighiLATA}, one can derive a circuit-like structure for a QCA, thereby extending the result of \cite{SchumacherWerner} to the $n$-dimensional case. We have also demonstrated that the constructive/operational definition of \cite{PerezCheung} can be given a rigorous axiomatics. And by doing these two things we have shown that the definitions of \cite{SchumacherWerner} and \cite{PerezCheung} are actually equivalent up to ancillary cells. This clearly reinforces the feeling that the community has now got a well-axiomatized and yet concrete definition of $n$-dimensional QCA.\\

\section{Perspectives} \label{pers}

There are many situations in physics where we want to study a unitary operator $U$ over a large Hilbert space $\mathcal{H}$, and struggle to obtain a practical representation for it. Often, however, these infinite dimensional Hilbert spaces arise from a position degree of freedom. By virtue of the principle according to which information travels at bounded speed, we can then think about ``cutting space into different pieces'' such that at each time step, the state of a piece depends solely on that of its neighbours. \nonqip{Whenever this happens Theorem \ref{th:locrep} applies and lets you write $U$ as $(\bigotimes D^\dagger)(\bigotimes K_x)(\bigotimes E)$, where $D$ and $E$ are local to each piece and $K_x$ is local to piece $x$ and its neighbours.}\qip{Whenever this happens our main theorem applies, which lets you decompose this $U$ as smaller operations, each of them well-localized upon a neighbourhood only.}\\
And so this is saying something very general which can be summarized by ``Unitarity plus causality implies localizability''. If a global evolution is locally implementable, this means we can focus on understanding local interactions between physical elements, and then the global evolution will just turn out to be a composition of them. And so this statement bridges a certain gap between general physical principles and the study of elementary interactions. \nonqip{Unfortunately there}\qip{There} were two \nonqip{strongly limiting}\qip{important} assumptions underlying this result, which we would be interested to lift.\\
The first one is unitarity. A not so uncommon belief amongst theoretical physicists is that the universe being a closed system it should evolve unitarily. Nevertheless this is clearly an unpractical view --- any everyday physical system is an open system, noisy due to its interactions with the outside world, amongst which any measurement we may wish perform upon the system. Hence one of our most wanted open problem at the moment is to extend \qip{the structure theorem}\nonqip{Theorem \ref{th:locrep}} to quantum operations. Clearly the way causality was axiomatized here will not entail localizability in an open systems setting (cf. PR-Boxes etc. \cite{Beckman}), so part of the challenge is to come up with a reinforced, yet intuitive notion of causality. \nonqip{Several cases of QCA under noise have been studied in \cite{KonnoSTAT, KonnoMMENT, BrennenWilliams}, which might perhaps be a guidance.}\\
The second one is discreteness. Clearly a point which is also rather open to discussion is whether we are indeed allowed to divide up the universe into different ``places'' whose state depend only on that of the closest neighbour, in the sort of abrupt and discrete manner which we use here. Hence we would like to study the relationship between QCA and continuous-space continuous-time models, maybe building upon what has been done for quantum simulation from Quantum Lattice Gas Automata\nonqip{ \cite{MeyerQLGI, MeyerQLGII, BoghosianTaylor1, BoghosianTaylor2, Vlasov, LoveBoghosian}}.\\
More generally it is our intention to understand the extent in which ``causality implies localizability'' could be made into a general principle. Such a principle would impact theoretical physics, by providing operational descriptions of global evolutions in physics. But it would also impact theoretical computer science, as these operational descriptions become closer and closer to being computable descriptions of global evolutions in physics.

\nonqip{
\section{Acknowledgements} We would like to acknowledge Gilles Dowek, Mehdi Mhalla, Renan Fargetton, Torsten Franz, Holger Vogts, Jarkko Kari, J\'er\^ome Durand-Lose, Philippe Jorrand.
}

\bibliography{../Bibliography/biblio}
\bibliographystyle{plain}

\end{document}